\documentclass[runningheads]{llncs}

\usepackage{eccv}

\usepackage{eccvabbrv}

\usepackage{graphicx}
\usepackage{booktabs}
\usepackage{algorithm}
\usepackage{amsmath} 
\usepackage[accsupp]{axessibility}  
\usepackage{wrapfig}

\usepackage[pagebackref,breaklinks,colorlinks,citecolor=eccvblue]{hyperref}

\usepackage{orcidlink}
\usepackage{microtype}
\usepackage{epsfig}
\usepackage{amsmath}
\usepackage{amssymb}
\usepackage{booktabs}
\usepackage{algorithmic}
\usepackage{mathtools}
\usepackage{multirow}
\usepackage{multicol}
\usepackage{makecell}

\begin{document}
\title{SSL-Cleanse: Trojan Detection and Mitigation in Self-Supervised Learning} 


\author{Mengxin Zheng\inst{12*}\and
Jiaqi Xue\inst{1*} \and
Zihao Wang\inst{2}\and
Xun Chen\inst{3} \\ 
Qian Lou\inst{1} \and Lei Jiang\inst{2}\and
Xiaofeng Wang \inst{2}}

\authorrunning{M. Zheng et al.}

\institute{University of Central Florida, Orlando, Florida, USA\and
Indiana University Bloomington, Bloomington, Indiana, USA\and 
Samsung Research America, Mountain View, California, USA}

\def\thefootnote{$*$}\footnotetext{These authors contributed equally to this work. \\Corresponding Author Email: mengxin.zheng@ucf.edu.}

\maketitle

\begin{abstract}
Self-supervised learning (SSL) is a prevalent approach for encoding data representations. Using a pre-trained SSL image encoder and subsequently training a downstream classifier, impressive performance can be achieved on various tasks with very little labeled data. The growing adoption of SSL has led to an increase in security research on SSL encoders and associated Trojan attacks. Trojan attacks embedded in SSL encoders can operate covertly, spreading across multiple users and devices.
The presence of backdoor behavior in Trojaned encoders can inadvertently be inherited by downstream classifiers, making it even more difficult to detect and mitigate the threat. Although current Trojan detection methods in supervised learning can potentially safeguard SSL downstream classifiers, identifying and addressing triggers in the SSL encoder before its widespread dissemination is a challenging task.
This challenge arises because downstream tasks might be unknown, dataset labels may be unavailable, and the original unlabeled training dataset might be inaccessible during Trojan detection in SSL encoders. We introduce \textbf{SSL-Cleanse} as a solution to identify and mitigate backdoor threats in SSL encoders. We evaluated SSL-Cleanse on various datasets using 1200 encoders, achieving an average detection success rate of $82.2\%$ on ImageNet-100. After mitigating backdoors, on average, backdoored encoders achieve $0.3\%$ attack success rate without great accuracy loss, proving the effectiveness of SSL-Cleanse. The source code of SSL-Cleanse is available at \url{https://github.com/UCF-ML-Research/SSL-Cleanse}.

\end{abstract}

\section{Introduction}
Self-supervised learning (SSL) has seen remarkable advancements,
particularly in computer vision applications~\cite{chen2020simple,krishnan2022self,liu2022graph,chen2020mocov2}. This is particularly evident when labeled examples are scarce. Unlike supervised learning, SSL sidesteps the labor-intensive labeling process, training on pretext tasks generalizable to many downstream tasks~\cite{chen2020mocov2, grill2020byol,chen2021exploring}. 
Several studies have demonstrated that SSL can achieve comparable~\cite{grill2020byol} and in some cases even superior, performance in few-shot learning~\cite{wu2020self,yaman2021zero}. 
The extensive use of SSL has spurred security research and vulnerability exploration in SSL encoders, as evidenced by the emergence of various Trojan attacks~\cite{sslbackdoor, jia2021badencoder, zhang2022corruptencoder, xue2022estas,CTRL,liu2022poisonedencoder,zheng2023trojvit, zheng2023trojfair}.

\begin{figure}[t!]
\vspace{-0.1in}
  \centering
   \includegraphics[width=0.8\linewidth]{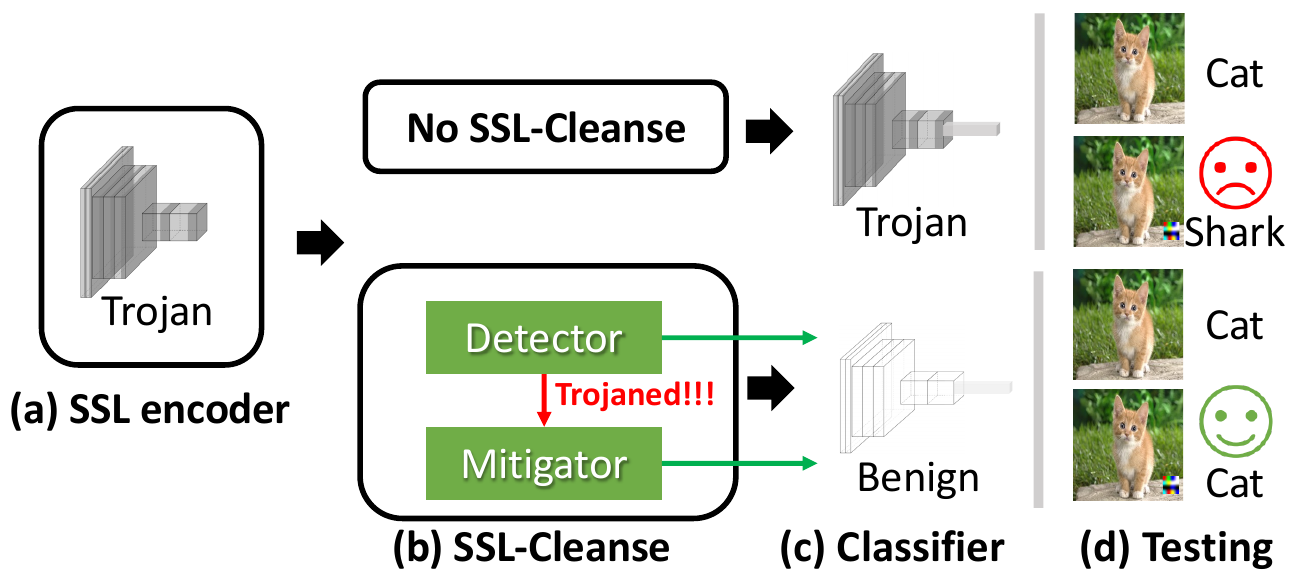}
   \caption{The overview of SSL-Cleanse. SSL-Cleanse has two components, Detector and Mitigator, aiming to remove the malicious behavior of Trojaned SSL encoders. 
   }
   \label{fig:overview}
   \vspace{-0.15in}
\end{figure} 

Malicious backdoor (a.k.a, Trojan) attacks~\cite{jia2021badencoder, loutrojtext, al2023trojbits,CTRL,sslbackdoor} in inputs, making the compromised model classify them into a predefined target class with high confidence.
If the trigger is removed from the input, the backdoored model will still exhibit normal behaviors with almost the same accuracy as its clean counterpart. One direction of SSL backdoor attacks assumes the attacker can control the training phase and modify the loss function to achieve training-control SSL backdoor~\cite{jia2021badencoder, xue2022estas} with high attack effects. In contrast, another popular SSL backdoor direction is training-agnostic~\cite{sslbackdoor, CTRL,zheng2023trojvit} where SSL backdoor attacks are executed in three phases. The first stage involves poisoning unlabeled datasets by adding triggers into a small fraction of target-class images. The second phase entails training the SSL encoder on the poisoned dataset to establish a connection between trigger and target-class images. In the final step, any downstream classifiers that are fine-tuned on the backdoored encoder inherit the backdoor behavior. The current training-agnostic backdoor attacks have demonstrated an attack success rate of over 98\% on the ImageNet-100 dataset~\cite{CTRL}. Our goal is to scan SSL encoders and mitigate backdoor threats against such attacks.

Trojan attacks in SSL encoders~\cite{sslbackdoor, CTRL} are perilous not only lies in their competitive attack success rates but also their covert functionality and broad reach across users and devices.
Firstly, pre-trained SSL encoders are typically spread out in real-world scenarios and subsequently fine-tuned for downstream classifiers. However, these downstream classifiers may inadvertently inherit the backdoor behaviors of Trojaned encoders. While current popular Trojan detection techniques~\cite{neural_cleanse,liu2019abs,kolouri2020universal} in supervised learning may have the potential to protect SSL downstream classifiers, detecting and mitigating triggers in the SSL encoder prior to its wide distribution is a complex undertaking. Recent encoder scanning techniques, as mentioned in \cite{decree}, recognize their inability to detect these Trojaned SSL encoders~\cite{sslbackdoor, CTRL} because of the distinct covert attack characteristics. We contend that detecting and mitigating Trojans in SSL encoders is crucial since it can impede the malicious distribution of Trojaned encoders. However, there is a research gap to bridge the popular backdoor defense methods in supervised learning with an SSL encoder. Detecting Trojans in SSL encoders is challenging due to unknown downstream tasks, unavailable dataset labels, and limited access to the original training dataset.
Even the linear probe strategy, which builds downstream classifiers, fails to detect these threats, as elaborated in the \textit{Limitations of Related Backdoor Defense} section. So \textit{it is crucial to implement effective detection and defense against such backdoor attacks on SSL encoders}.

This paper introduces \textit{SSL-Cleanse}, a novel backdoor defense method as illustrated in Figure~\ref{fig:overview}. The proposed approach comprises two main components, namely Detector and Mitigator. The Detector is responsible for identifying the presence of Trojan in an SSL encoder, and if found, the Mitigator can mitigate the backdoor attack effect. Our SSL-Cleanse overcomes the challenges of backdoor detection without knowing the labeled data and the downstream tasks. 

Our contributions can be summarized as follows:
\begin{itemize}
\item 
We design a framework to detect training-agnostic attacks in SSL encoders without downstream labels. We reveal it is possible to prevent the dissemination of Trojaned SSL encoders via our SSL-Cleanse. 
\item We introduce the Sliding Window Kneedle (SWK) algorithm to auto-estimate cluster counts in unlabeled datasets, aiding representation clustering. We also present the representation-oriented trigger reverse (ROTR) method for SSL trigger inversion, alongside the Self-supervised Clustering Unlearning (SCU) algorithm to mitigate SSL encoder backdoors.

\item We validate the effectiveness of the proposed SSL-Cleanse with extensive experiments on 1200 encoders. 
\end{itemize}

\section{Background and Related Works}\label{background}

\textbf{Self-Supervised Learning.}
Leveraging the large unlabeled data available in the real world is essential.
Self-Supervised Learning (SSL) is the most popular method to learn representations from complex unlabeled data~\cite{chen2020simple,krishnan2022self}.  Pre-training an encoder with significant unlabeled data and fine-tuning it with a small amount of labeled data has been shown to achieve comparable performance to using large labeled datasets with supervised learning methods for various downstream tasks~\cite{chen2020simple,krishnan2022self,liu2022graph,chen2020mocov2,jaiswal2020survey,lan2019albert,he2020momentum}. Furthermore, SSL techniques that rely on instance discrimination, such as SimCLR~\cite{chen2020simple} and MoCo V2~\cite{chen2020mocov2}, have become increasingly popular for learning competitive visual representations through the use of contrastive loss. 
Typically, an SSL classification task involves pre-training an image encoder, constructing a classifier, and subsequent fine-tuning.

\begin{table}[h!]
\centering
\scriptsize
\setlength{\tabcolsep}{3pt}
\caption{Limitations of current backdoor detectors in assessing SSL encoders, such as SSL-Backdoor~\cite{sslbackdoor} and CTRL~\cite{CTRL}. The encoder undergoes pre-training on CIFAR-10, while the built classifiers are evaluated on CIFAR-10, STL-10, and GTSRB. These datasets have labels that overlap, partially overlap, or do not overlap with the encoder. If NC~\cite{neural_cleanse} Index $>$ 2.0 and ABS~\cite{liu2019abs} REASR $>$ 0.88, the model is seen as Trojaned. }
\begin{tabular}{cccc}\toprule
\multirow{2}{*}{\makecell[c]{SSL Attack\\Method}} & \multirow{2}{*}{\makecell[c]{Downstream Task\\(Linear probe)}} & NC & ABS \\\cmidrule{3-4}
& & Anomaly Index & REASR \\\midrule
\multirow{3}{*}{\makecell[c]{SSL-\\Backdoor}} & CIFAR-10 & $2.05$ & $0.89$ \\
& STL-10 & $1.42$ & $0.34$ \\
& GTSRB & $1.68$ & $0.29$ \\\midrule
\multirow{3}{*}{CTRL} & CIFAR-10 & $1.52$ & $0.52$ \\
& STL-10 & $1.28$ & $0.44$ \\
& GTSRB & $1.16$ & $0.37$ \\\bottomrule
\end{tabular}
\label{t:limitation}
\vspace{-0.2in}
\end{table}

\noindent\textbf{SSL Backdoor Attacks.} In SSL backdoor attacks, the first line of research~\cite{jia2021badencoder} links the trigger to the downstream-task label, necessitating triggers to be appended to varying class inputs. Compared to other directions, it assumes a stronger threat model in which the
adversary dominates the training process, e.g., modifying the loss functions. Another methodology~\cite{liu2022poisonedencoder,xue2022estas} focuses on specific input sets, formulating poisoned data by combining these inputs with their corresponding reference representations. However, it functions only with predetermined specific inputs and not with any inputs that contain embedded triggers. A distinct approach ~\cite{sslbackdoor, CTRL} avoids modifying the training phase, achieving a high attack success rate by merely connecting the trigger with desired unlabeled samples. Compared to the first approach, these SSL attacks are much more challenging to discern since triggers are solely attached to the unlabeled targets, and combining triggers with other targets does not consistently yield similar representations~\cite{decree}. \textit{In our study, we aim to detect these third-category attacks.} 

\noindent\textbf{Limitations of Related Backdoor Defense.}
Prior scanners like NC~\cite{neural_cleanse} and ABS~\cite{liu2019abs} face challenges in detecting backdoors in SSL encoders.
The primary reasons include the often-unknown downstream tasks/labels. While the linear probe method, i.e., constructing downstream classifiers from various datasets using pre-trained encoders, offers an alternative, it is not efficacious in scanning SSL encoders. For example, we built a backdoored encoder using CIFAR-10~\cite{cifar10} with a specific label, \textit{airplane}, as the target. This encoder was then utilized to train three distinct downstream classifiers on CIFAR-10, STL-10~\cite{stl10}, and GTSRB~\cite{GTSRB}. The results from applying NC and ABS to these classifiers are shown in Table~\ref{t:limitation}. For scenarios where the encoder and the downstream task share the same dataset, both NC and ABS can detect Trojaned classifiers and hence the backdoored encoders for SSL-Backdoor~\cite{sslbackdoor}, with encoder's Anomaly Index of 2.05 $>$ 2.00 in NC and a REASR of 0.89 $>$ 0.88 in ABS. Yet, the detection capability requires the knowledge of the downstream tasks and the  detection capability diminishes when the datasets (like STL-10 or GTSRB) either only partially overlap or do not align at all with CIFAR-10. For the CTRL instance, both tools fail in detection, even when there's label congruence between the encoder and classifiers. The introduction of global frequency triggers in CTRL exacerbates the detection difficulty. 
While the recent study DECREE~\cite{decree} effectively detects backdoors in training-controlled attackers, like BadEncoder~\cite{jia2021badencoder}, it acknowledges the failure in identifying stealthy training-agnostic attacks~\cite{sslbackdoor, CTRL}, e.g., $\sim$ 50\% detection accuracy. Also, scanners such as PatchSearch~\cite{tejankar2023defending} and ASSET~\cite{pan2023asset} are orthogonal to our method since they are introduced to identify poisoned samples in a training dataset. SSL-ABD~\cite{yang2023ssl} employs adversarial simulation of trigger patterns and subsequent removal of the backdoor from the compromised encoder through feature embedding distillation. However, as the defender lacks knowledge of the target class, simulating the trigger pattern becomes challenging, hindering effective backdoor removal. In our threat model, the SSL encoder is required but there's no requirement for the poisoned training dataset.

\begin{figure*}[t!]
  \centering
   \includegraphics[width=\linewidth]{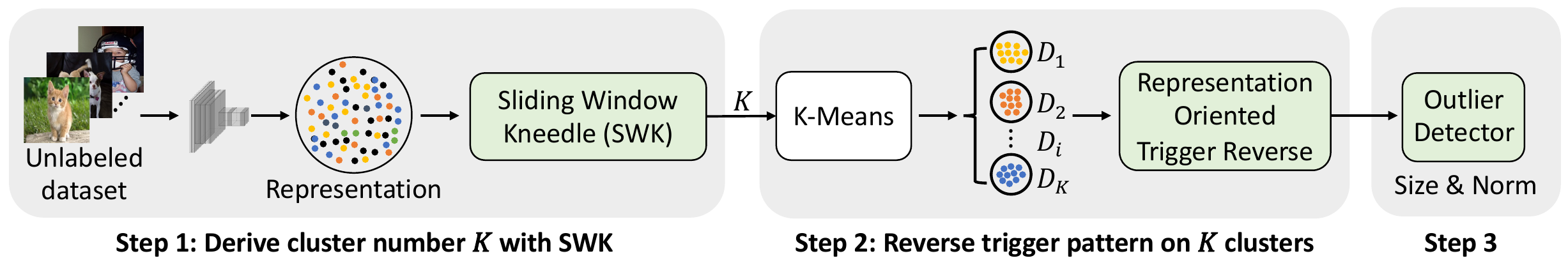}
   \caption{The workflow of SSL-Cleanse detector. Step 1: Unlabeled data samples are processed through the SSL encoder to compute their representations. The SWK algorithm is then utilized to process representations and determine the number of clusters. Step 2: Using K-Means with the derived cluster number (\textit{K}) and representation, \textit{K} clusters are established. Then, the Representation Oriented Trigger Reverse algorithm is employed to generate \textit{K} trigger patterns.
   Step 3: Accessing if any of the \textit{K} triggers are outliers in terms of their size and norm. The identified outlier indicates the encoder is Trojaned.}
   \vspace{-0.1in}
   \label{fig:workflow}
\end{figure*}

\section{SSL-Cleanse Design Overview}\label{setting}

\textbf{Defense Assumptions and Goals.}
We assume that the defender has access to a pre-trained SSL encoder, a small portion of the unlabeled dataset (which could be distinct from the training set, i.e., SSL-Cleanse does not require attackers to disclose their poisoned dataset). 

\noindent\textbf{Goals.} We have two specific goals:
\begin{itemize}
\item \textbf{Detecting backdoor:} Our aim is to make a binary determination regarding the potential backdoor infection of a given SSL encoder. If infected, we also want to identify the potential target classes of the backdoor attack. 
 

\item \textbf{Mitigating backdoor:} We plan to reversely generate the trigger used by the attack. Our ultimate goal is to deactivate the backdoor, making it ineffective. This requires eliminating the backdoor while preserving the classification performance of the SSL encoder for normal inputs. 
\end{itemize}

\noindent\textbf{Challenges and Motivation.} An inherent challenge of SSL backdoor detection lies in the indeterminate nature of the target class. While this can be addressed by iteratively searching through all classes, this method is not suitable for SSL encoders due to the uncertainty in class numbers for unlabeled datasets in SSL. Traditional clustering methods, like K-Means~\cite{kmeans}, necessitate manual predetermination of cluster counts $K$. We observe that though automatic cluster number determination methods, such as the Kneedle algorithm~\cite{kneedle} exist, they struggle with accurate class number prediction for SSL encoders. For instance, the Kneedle algorithm identifies only 20 classes within the ImageNet-100 dataset. This discrepancy could arise from noisy samples affecting the clustering of SSL representations. Other methods like DBSCAN~\cite{campello2015hierarchical} and Mean Shift~\cite{ghassabeh2015sufficient} rely on predefined parameters like density radius and kernel width. Determining optimal values for these parameters presents a challenge. To enhance cluster number prediction accuracy, we propose the Sliding-Window Kneedle (SWK) algorithm, which addresses noise influence by averaging adjacent Kneedle scores within a sliding window. And this approach is adaptable across various clustering methods.

Another challenge is how to efficiently generate triggers and identify outlier triggers given predicted cluster numbers, encoder, and sampled unlabeled dataset.  This challenge is exaggerated when we target to detect attacks~\cite{CTRL} based on frequency-domain global triggers since one cannot identify outlier triggers according to the trigger size. 
For these reasons, we are motivated to propose Representation-Oriented Trigger Reverse (ROTR) method to generate triggers via clustered representations. For identifying patch-wise trigger SSL attacks, specifically SSL-Backdoor as referenced in ~\cite{sslbackdoor}, we continue to rely on trigger size as a key outlier detection metric. However, when detecting the frequency-domain SSL attack, CTRL as cited in ~\cite{CTRL}, we propose to use the magnitude of the trigger as the pivotal criterion. These detection methodologies for both attack types are unified under RORP. Upon identifying the triggers, it becomes feasible to pinpoint their related target-class clusters. Following this, we introduce the Self-supervised Clustering Unlearning (SCU) approach to mitigate backdoor threats in SSL encoders, resulting in a cleansed encoder. 

\section{SSL-Cleanse Detector}  \label{approach}

\textbf{Workflow.}
\label{sec:Detection}
The workflow of the SSL-Cleanse detector, as shown in Figure~\ref{fig:workflow}, consists of three steps. First, the SSL encoder processes a small amount of unlabeled data to generate representations. Using these, the Sliding Window Kneedle (SWK) algorithm determines the cluster count $K$. Subsequently, K-Means creates \textit{K} clusters, from which the Representation Oriented Trigger Reverse algorithm derives \textit{K} trigger patterns. Finally, if any trigger significantly deviates in size or magnitude/norm, the encoder is deemed Trojaned. The details of each step and the associated algorithms are elaborated below.  

\begin{figure}[h!]
\centering
\includegraphics[width=0.75\linewidth]{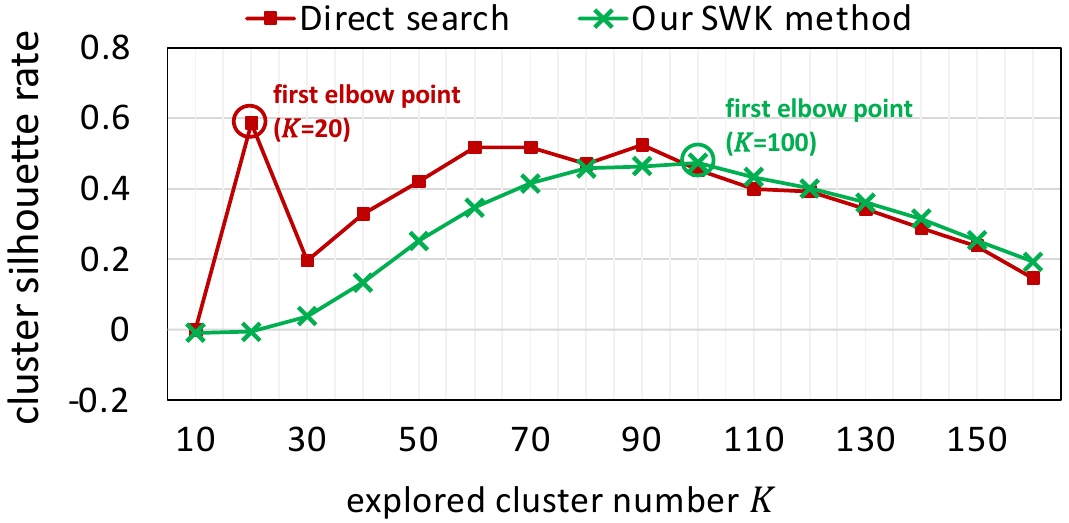}
\caption{Comparison of our SWK method and direct search (Kneedle) method on ImageNet-100 dataset. Our SWK method yields more stable and accurate K.}
\label{fig:SWK}
\vspace{-0.2in}
\end{figure}

\noindent\textbf{Sliding Window Kneedle.} 
Class numbers for unlabeled training samples in SSL encoders are often unknown, even to model developers. 

At first glance, clustering appears as an intuitive approach to determining class numbers. Yet, conventional methods such as K-Means~\cite{kmeans} demand a predetermined value for $K$. Opting for an automated determination of the cluster number $K$ offers a more flexible solution. To this end, we initially investigated existing automatic methods, with a focus on the Kneedle algorithm~\cite{kneedle}.

Given the SSL samples $D$ and encoder $f$, the Kneedle algorithm starts by initializing a list with potential $K$ values to examine. For each $K$ in this list, the K-Means method clusters the representation $f(D)$, producing a clustering outcome. The silhouette score for this clustering ranges from -1 to 1. A high value signifies that the data point is well-suited to its own cluster and has a poor fit with neighboring clusters.

The silhouette score for a particular data point is computed using: $ s(i) = \frac{b(i) - a(i)}{\max\{a(i), b(i)\}} $  where
$s(i)$ represents the silhouette coefficient for data point $i$,
$a(i)$ denotes the mean distance between the $i^{th}$ data point and other points within the same cluster,
$b(i)$ is the smallest mean distance between the $i^{th}$ data point and points in a different cluster, minimized over all clusters. The clusters' overall silhouette score is the mean silhouette coefficient of all instances.

The Knee/elbow point on the plot, which represents silhouette scores over varying $K$ values, indicates the optimal position.
To identify the Knee point, one should normalize the silhouette curve, adjust the origin to $[0,0]$ and ensure the endpoint aligns with $[1,1]$.
Next, measure the distance of each point on the curve from the direct line connecting the origin $[0,0]$ to the endpoint $[1,1]$.
The point that is furthest from this direct line is recognized as the knee point of the curve.

\begin{algorithm}
\caption{Sliding Window Kneedle for SSL Cluster Num.}
   \label{alg:SWK}
\begin{algorithmic}
    \STATE {\bfseries Input:} SSL samples $D$, encoder $f$, pre-defined K\_list
    \STATE {\bfseries Output:} predicted cluster number $K$ 
    \STATE initialize clusters\_list, s\_list, padded\_s\_list, d\_list = []
    \FOR{$i=0$ {\bfseries to} $len(K\_list)$}
    \STATE clusters\_list.$append(k\-means(f(D)$ ,K\_list[$i$]))
    \STATE s\_list$.append(silhouette(f(D)$,clusters\_list[$i$]))
    \ENDFOR
    \STATE initialize window size $w$ as a small odd number, e.g., 3
    \STATE initialize swk\_s\_list to zero values of s\_list's structure
    \STATE padded\_s\_list $ \gets$ pad $\frac{w-1}{2}$ zeros to head and tail of $s\_list$
    \FOR{$i=1$ {\bfseries to} $len$(s\_list)}
    \STATE swk\_s\_list[i]= $\frac{1}{w}\sum_{j=0}^w $ padded\_s\_list[i+j]
     \STATE d\_list.$append$($norm$(swk\_s\_list[i])-$norm$(K\_list))
    \ENDFOR    
    \STATE $K \gets$ index of maximum entry in (d\_list)
\end{algorithmic}
\end{algorithm}

In Figure~\ref{fig:SWK}, we illustrate the silhouette curvature distance for each point $K$. The line corresponding to the direct search showcases the application of the Kneedle algorithm. The most considerable distance is observed when $K=20$, indicating the elbow point of the Silhouette score is 20. Nonetheless, this estimation might not be precise, given that the SSL encoder 
$f$ is trained on the ImageNet-100 dataset. This discrepancy may arise from the impact of noisy samples on the clustering of SSL representations.

To address the high-dimensional noise and varied representations in the encoder's outputs, we enhanced the direct search (Kneedle) approach by introducing a sliding window technique with a window size of 
$w$. The underlying idea is to compute the average silhouette scores for neighboring \(K\) values, aiming to refine the silhouette curvature. As depicted in Algorithm \ref{alg:SWK}, our new algorithm, Sliding Window Kneedle (SWK), ingests the encoder $f$, SSL samples $D$, and a pre-set K\_list. The output is the anticipated cluster count, $K$. The foundation of SWK relies on the Kneedle algorithm's methodology to determine K-Means clustering clusters\_list[i] and silhouette scores for every specified K\_list[i]. Post silhouette score list (s\_list) computation, SWK introduces a window dimension to derive the mean silhouette score, symbolized as swk\_s\_list, for the neighboring $w$ scores. This process necessitates padding zeros to s\_list, resulting in padded\_s\_list. The silhouette curvature distance, d\_list[i], is deduced from the disparity between the normalized swk\_s\_list and K\_list, with normalization between [0,1]. The position of the largest value in d\_list dictates the predicted 
$K$. Figure~\ref{fig:SWK} illustrates that the evolved SWK curve offers better clarity in pinpointing the elbow juncture, diminishes noise, and assures a more precise cluster estimation.

\begin{algorithm}[htb!]
\caption{Representation-Oriented Trigger Reverse}
   \label{alg:ROTR}
\begin{algorithmic}
    \STATE {\bfseries Input:} clustered samples $D_{i \in [1,K] }$, encoder $f$, cluster number $K$, epoch number $E$
    \STATE {\bfseries Output:} masks $m^1_i$,$m^2_i$ perturbation $\Delta^1_i$,$\Delta^2_i$ of $K$ clusters
    \FOR{$i=1$ {\bfseries to} $K$}
    \STATE initialize masks $m^1_i$,$m^2_i$ and perturbation $\Delta^1_i$,$\Delta^2_i$
    \FOR{$e=1$ {\bfseries to} $E$}
    \STATE $x_i \gets$ randomly sample an image from $D_i$ \# target
    \STATE $x_j \gets$ randomly sample an image from $D_{j\neq i}$
    \STATE $x^1_j \gets (1-m_i^1) \cdot x_j+m^1_i\cdot \Delta^1_i$
    \STATE $x^2_j \gets (1-m_i^2) \cdot x_j+m^2_i\cdot \Delta^2_i$
    \STATE $loss_{size} \gets \mathcal{L}^{size}_{MSE}(f(x_i), f(x_j^1))$
    \STATE $loss_{norm} \gets \mathcal{L}^{norm}_{MSE}(f(x_i), f(x_j^2))$
    \STATE $m^1_i, \Delta^1_i \gets update(m^1_i, \Delta^1_i, loss_{size})$
    \STATE $m^2_i, \Delta^2_i \gets update(m^2_i, \Delta^2_i, loss_{norm})$
    \ENDFOR
    \ENDFOR
\end{algorithmic}
\end{algorithm}

\noindent\textbf{Representation-Oriented Trigger Reverse.}
\label{sec:Detection:loss}
Once the cluster count $K$ is determined, K-Means can be readily employed to produce $K$ clustered samples, denoted by $D_{i \in [1,K] }$, using the provided $D$. The subsequent goal is to backtrack and identify Trojan triggers that are either compact in size or have a minimal $l_1$-norm magnitude that can mimic the representations of target-class samples.
Algorithm~\ref{alg:ROTR} presents the ROTR approach. The core concept involves creating two triggers for each cluster representation. It then determines which representation clusters can yield outlier triggers, either having smaller patch-based trigger sizes \(|m_i^1|\) for patch-based trigger attacks or lesser trigger magnitudes \(m_i^2 \cdot \Delta_i^2\) for global-trigger attacks. Initially, we designate trigger masks \( m_i \) to specify the location of the trigger pixel, while \( \Delta_i \) signifies the associated pixel value for the \( i \)-th trigger. Consequently, the trigger \( r_i \) is computed as \( m_i \cdot \Delta_i \). To identify both patch-based triggers and frequency-domain global triggers, we suggest initiating two distinct trigger sets: \( r_i^1 = m_i^1 \cdot \Delta_i^1 \) and \( r_i^2 = m_i^2 \cdot \Delta_i^2 \), where \( m_i^1 \) and \( m_i^2 \) represent masks and \( \Delta_i^1 \) and \( \Delta_i^2 \) delineate the pixel values of the triggers. For each cluster, we conduct \( E \) epochs to produce these twin trigger sets. Specifically, we randomly select an image from \( D_i \) to as the clean sample \( x_i \) and subsequently get sample to add trigger, termed \( S_j \), from \( D_{j \neq i} \). We then affix the pre-established two trigger sets to \( x_j \), resulting in \( x_j^1 \) and \( x_j^2 \) respectively.

The clean image $x_i$ and images with trigger $x_j^1$ and $x_j^2$  are then sent to an encoder $f$, and two separate loss functions are employed to update the trigger such that its representation can have more similarity with $x_i$'s feature. The loss functions used in the ROTR to optimize $m_i^1$, $m_i^2$ and $\Delta_i^1$, $\Delta_i^2$ on the $i^{th}$ cluster are given by Equation~\ref{e:lsize} and Equation~\ref{e:lnorm}, respectively. Each loss function comprises two components. The foremost objective of the first term is to guarantee the similarity between the image patched with the trigger and the target class image in the feature space. The second term is responsible for constraining the size or the trigger magnitude/norm of the reversed trigger, we adapted $\lambda$ dynamically during optimization in our experiment. The dynamic scheduler is described in the supplementary material.  Finally, the resulting trigger patterns are sent to an outlier detector for determination.

{\footnotesize
\begin{equation}
    \label{e:lsize}
        \mathcal{L}^{size}_{MSE}(f(x_i), f(x_j^1))=-\frac{<f(x_i), f(x_j^1)>}{||f(x_i)|| \cdot ||f(x_j^1)||} + \lambda \cdot |m^1_i| 
\end{equation}
}

{\footnotesize
\begin{equation}
    \label{e:lnorm}
        \mathcal{L}^{norm}_{MSE}(f(x_i), f(x_j^2))=-\frac{<f(x_i), f(x_j^2)>}{||f(x_i)|| \cdot ||f(x_j^2)||} + \lambda \cdot |m^2_i\cdot \Delta^2_i| 
\end{equation}
}

\noindent\textbf{Size-Norm Trigger Outlier Detector.}
Beyond just using trigger size to identify outliers for patch-wise trigger attacks, e.g., SSL-Backdoor~\cite{sslbackdoor}, we recognize trigger magnitude/norm as a key criterion for global-trigger attacks stemming from frequency-domain disturbances, e.g., CTRL~\cite{CTRL}. Our proposed Size-Norm Trigger Outlier Detection (STOD) is designed to detect both patch-based and global-based triggers. In particular, given a trigger size list $[(m_1^1, \Delta_1^1), ..., (m_K^1,\Delta_K^1)]$ and a trigger norm list $[(m_1^2, \Delta_1^2), ..., (m_K^2,\Delta_K^2)]$, the STOD outputs the detection result, i.e., benign or Trojaned with a trigger dictionary $t_s$. For $K$ cluster, we iteratively check if the trigger size $|m^1_i|$ and trigger norm $|m_i^2\cdot \Delta_i^2|$ is outlier in the list of $|m_{[1:K]}^1|$  and $|m_{[1:K]}^2\cdot \Delta_{[1:K]}^2|$ using the function \textit{is\_outlier($x_i, \boldsymbol{x}$)}. Here \textit{$is\_outlier(x_i, \boldsymbol{x})$} returns True if $M(x_i, 
\boldsymbol{x})>2$, otherwise False; The Anomaly Index function $M(x_i, \boldsymbol{x})=\frac{|x_i-median(\boldsymbol{x})|}{c\cdot median(|x-median(\boldsymbol{x})|)}$ is used to ascertain if $x_i$ is an anomaly. $c$ is a constant estimator which equals $1.4826$. If we discover a trigger that is substantially smaller than the other candidates, we classify it as an outlier and store its cluster number $i$ and itself in a dictionary $t_s$, i.e., $t_s[i]=m_i\cdot \Delta_i$. If $t_s$ remains empty, we conclude that the encoder is benign. However, if $t_s$ contains any triggers, we classify the encoder as a Trojaned encoder.

\begin{figure}[ht!]
\vspace{-0.15in}
  \centering
\includegraphics[width=0.8\linewidth]{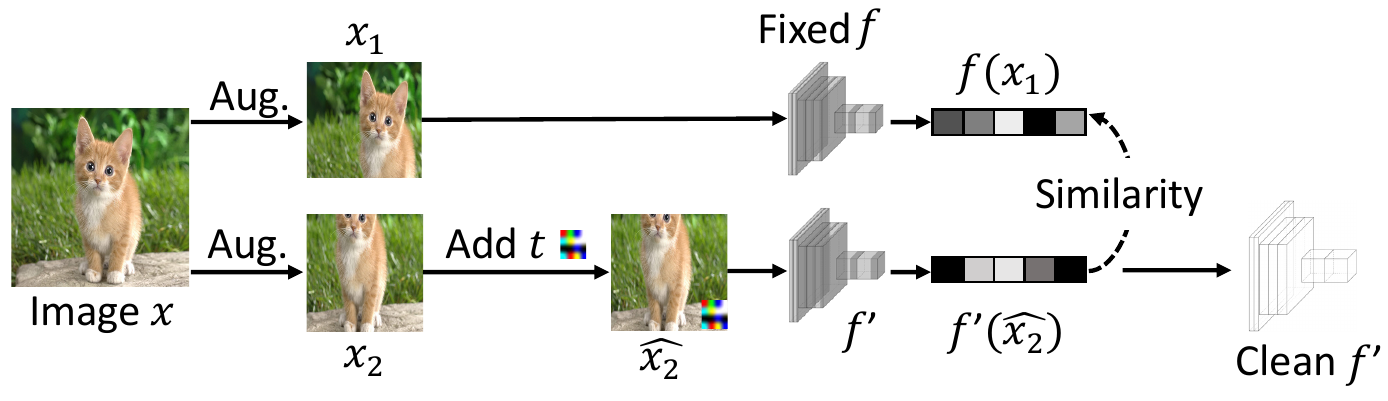}
\vspace{-0.1in}
   \caption{Illustration of Self-supervised Clustering Unlearning (SCU). The image \( x \) is sampled from a cluster distinct from the cluster producing trigger \( t \).}
     \label{fig:mitigation}
     \vspace{-0.2in}
\end{figure}

\section{SSL-Cleanse Mitigator} \label{sec:mitigation} 
Once we have the triggers $t_s$ generated by our detector, their corresponding target-class clusters become identifiable. Subsequently, we introduce a Self-supervised Clustering Unlearning (SCU) strategy to mitigate SSL encoder backdoor threats, leading to a purified encoder. The mitigation approach is detailed in Algorithm~\ref{alg:mitigation}. This method accepts cluster samples \( D_{i \in [1, k]} \), a Trojaned encoder \( f \), and the trigger list \( t_s \) as inputs, producing a purified encoder \( f' \) in return. Since $K$ clusters are generated, we need to clean them one by one. For each cluster $i$, we iteratively select clean image $x$ from each cluster samples $D_i$, and augment the image to create a new training sample consisting of the augmented images $x_{1}$ and $x_{2}$. Subsequently, we randomly select a trigger from $t_s$ excluding $t_s[i]$ to make sure the image $x$ is sampled from a cluster distinct from the cluster producing trigger $t$. Then we attach trigger $t$ to $x_2$ or directly use $x_{i2}$ without adding a trigger. The probability is $50\%$ which is the meaning of $equalSample$ in Algorithm~\ref{alg:mitigation}. The trigger insertion rate is critical for balancing attack mitigation with the preservation of clean accuracy. A $50\%$ rate is shown to be optimal—it sufficiently counters attacks without compromising clean accuracy. Notice that here we pass these new training samples through the Trojaned encoder $f$ to obtain their respective representations. We then optimize the similarity between the representations using a loss function by fixing the model $f$ and updating the encoder $f'$ to eliminate the Trojan trigger effects, resulting in a clean encoder. Figure~\ref{fig:mitigation} illustrates the mitigation process, where a clean image of a cat $x$ is augmented to generate two cat images $x_{1}$ and $x_{2}$, one that is the same and another that has a 50\% chance of being attached with a Trojan trigger. The clean image is sent to a reference model $f$, which maintains its parameters and representations unchanged. In contrast, the other model $f'$ is updated with the new training sample to remove the Trojan trigger effect. The updated encoder is deemed to be clean after it is able to accurately classify data, even in the presence of the Trojan trigger.


\begin{algorithm}[t!]
   \caption{Self-supervised Clustering Unlearning}
   \label{alg:mitigation}
\begin{algorithmic}
    \STATE {\bfseries Input:} $D_{i \in [1, k]}$, Trojaned encoder $f$, trigger list $t_s$
    \STATE {\bfseries Output:} a clean encoder $f^{\prime}$
    \STATE Initialize $f^{\prime} \gets f$
    \FOR{$i=1$ {\bfseries to} $K$}
    \FOR{$x$ in $D_i$} 
    \STATE $x_1 \gets aug_1(x)$; $x_2 \gets aug_2(x)$ 
    \STATE $t \gets$ randomly selected from $t_s$ excluding $t_s[i]$
    \STATE $\hat{x_2} \gets equalSample\{aug_2(x\oplus t), aug_2(x)\}$
    \STATE $z, z^{\prime}=f(x_1), f^{\prime}(\hat{x_2})$
    \STATE $loss \gets -similarity(z, z^{\prime})$
    \STATE $f^{\prime} \gets update(f, loss)$
    \ENDFOR
    \ENDFOR
\end{algorithmic}
\end{algorithm}

\section{Experimental Methodology} \label{method}
\textbf{Dataset}. Our experiments were conducted on benchmark datasets: CIFAR-10~\cite{cifar10} and ImageNet-100~\cite{imagenet}. 
ImageNet-100 is a random subset of 100 classes from the larger ImageNet dataset and contains around 127,000 training images, which is widely used in prior SSL attacks~\cite{sslbackdoor,CTRL}.

\noindent\textbf{SSL Attacks and Encoders}. To assess the effectiveness of our detector against various attack methods, we evaluated it against two backdoor attacks, namely SSL-Backdoor~\cite{sslbackdoor}and CTRL~\cite{CTRL} over BYOL~\cite{grill2020byol}, SimCLR~\cite{chen2020simple}, and MoCo V2~\cite{chen2020mocov2}, respectively. We created 50 benign encoders and 50 Trojaned encoders for each backdoor attack setting, resulting in 1200 encoders. We follow the above attack's setting to use ResNet-18~\cite{resnet} as encoder architecture.  

\noindent\textbf{Experimental Settings}. Our experiments are performed on two Nvidia GeForce RTX-3090 GPUs, each with a memory capacity of 24 GB. For the detector, the initial value of $\lambda$ is set up as $0.01$. 
For detection and mitigation running overhead, the detection method with $10\%$ of the ImageNet-100 training data consumes roughly 20 minutes, and the mitigation process requires approximately 7 minutes.

\noindent\textbf{Evaluation Metrics}. We define the following evaluation metrics to study the efficiency and effectiveness of our SSL-Cleanse. Detection Accuracy (\textbf{DACC}) is detection accuracy which is the ratio of correctly identified encoder types (either Benign or Trojan) relative to the total count of encoders. 
Attack Success Rate (\textbf{ASR}) is defined as the ratio of images that contain the trigger and are misclassified as the target class, to the total number of evaluated images. Accuracy (\textbf{ACC}) is the percentage of input images without triggers classified into their corresponding correct classes. \textbf{TP} indicates the true positive count, referring to Trojaned encoder numbers detected by our detector. \textbf{FP} represents false positives, indicating clean encoders misclassified as Trojaned encoders by our detector.

\section{SSL-Cleanse Results}

\textbf{Detection.} In Table~\ref{t:detector}, we present the performance of our detector on two training-agnostic SSL attacks including SSL-Backdoor and CTRL on three SSL methods and two datasets.   
In total, our detection accuracy (DACC) across 1200 (600 Trojaned and 600 benign) encoders stands at \(81.33\%\), illustrating the efficacy of our backdoor detection capabilities. In particular, for the ImageNet-100 dataset, the DACC is $>77\%$ and the average DACC is $82.17\%$. For the CIFAR-10 dataset, the DACC is $>76\%$ and the average DACC is $80.5\%$. 

\begin{table}[ht!]
\centering
\vspace{-0.15in}
\scriptsize
\setlength{\tabcolsep}{2pt}
\caption{The detection performance of our SSL-Cleanse.}
\begin{tabular}{cccccccc}\toprule
\multirow{2}{*}{Dataset} & \multirow{2}{*}{Method} & \multicolumn{3}{c}{SSL-Backdoor} & \multicolumn{3}{c}{CTRL} \\\cmidrule(lr){3-5} \cmidrule(lr){6-8}
& & TP & FP & DACC(\%) & TP & FP & DACC(\%) \\\midrule
\multirow{3}{*}{CIFAR-10} & BYOL & $35$ & $5$ & $80$ & $36$ & $4$ & $82$\\
& SimCLR & $33$ & $4$ & $79$ & $39$ & $5$ & $84$\\
& MoCo V2 & $31$ & $5$ & $76$ & $37$ & $5$ & $82$\\\midrule
\multirow{3}{*}{ImageNet-100} & BYOL & $38$ & $8$ & $80$ & $43$ & $8$ & $85$ \\
& SimCLR & $34$ & $7$ & $77$ & $46$ & $8$ & $88$ \\
& MoCo V2 & $36$ & $7$ & $79$ & $42$ & $8$ & $84$ \\
\bottomrule
\end{tabular}
\label{t:detector}
\vspace{-0.1in}
\end{table}

Table~\ref{t:detector} further indicates that our SSL-Cleanse can detect not only the patch-based trigger SSL attack SSL-Backdoor but also the frequency-based SSL attack CTRL. This capability stems from our method's utilization of trigger size and magnitude for patch-based trigger detection in tandem with frequency-domain detection. In particular, For SSL-Backdoor detection, SSL-Clease achieves $78.5\%$ DACC on average for both CIFAR-10 and ImageNet-100 datasets. For CTRL, it obtains $84.17\%$ average DACC. Against the CTRL, our detector identifies 46 TP from 50 Trojaned SimCLR encoders on ImageNet and registers 8 FP among 50 clean SimCLR encoders. Our SSL-Cleanse consistently exhibits reliable detection performance across prevalent SSL training techniques such as BYOL, SimCLR, and MoCo V2. In particular, our detector registers an average DACC of \(81.75\%\), \(82\%\), and \(80.25\%\) for BYOL, SimCLR, and MoCo V2, respectively.

\begin{table*}[ht!]
\centering
\scriptsize
\setlength{\tabcolsep}{1pt}
\caption{The mitigation performance of our SSL-Cleanse.}
\begin{tabular}{cccccccccc}\toprule
\multirow{3}{*}{Dataset} & \multirow{3}{*}{Method} & \multicolumn{4}{c}{SSL-Backdoor} & \multicolumn{4}{c}{CTRL} \\\cmidrule(lr){3-6}\cmidrule(lr){7-10}
& & \multicolumn{2}{c}{Before mitigation} & \multicolumn{2}{c}{After mitigation} & \multicolumn{2}{c}{Before mitigation} & \multicolumn{2}{c}{After mitigation} \\\cmidrule(lr){3-4}\cmidrule(lr){5-6}\cmidrule(lr){7-8}\cmidrule(lr){9-10}
& & ACC(\%) & ASR(\%) & ACC(\%) & ASR(\%) & ACC(\%) & ASR(\%) & ACC(\%) & ASR(\%) \\\midrule
\multirow{3}{*}{CIFAR-10} & BYOL & $83.42$ & $48.32$ & $82.14$ & $1.14$ & $83.19$ & $60.47$ & $82.59$ & $1.96$\\
& SimCLR & $84.88$ & $42.19$ & $83.53$ & $0.58$ & $80.74$ & $81.84$ & $79.60$ & $1.15$ \\
& MoCo V2 & $81.02$ & $37.95$ & $80.16$ & $0.92$ & $81.42$ & $77.51$ & $80.03$ & $1.62$ \\\midrule

\multirow{3}{*}{ImageNet-100} & BYOL & $60.57$ & $33.21$ & $60.24$ & $0.14$ & $53.33$ & $45.10$ & $52.65$ & $0.35$ \\
& SimCLR & $60.18$ & $31.85$ & $58.58$ & $0.62$ & $52.90$ & $44.98$ & $51.04$ & $0.33$ \\
& MoCo V2 & $61.57$ & $35.06$ & $60.10$ & $0.17$ & $50.62$ & $35.72$ & $48.88$ & $0.17$ \\
\bottomrule
\end{tabular}
\label{t:mitigator}
\end{table*}

\noindent\textbf{Mitigation.} Table~\ref{t:mitigator} compares the attack success rate (ASR) and clean accuracy (ACC) before and after applying the mitigator against SSL-Backdoor and CTRL on the CIFAR-10 and ImageNet-100 dataset. On average, the ASR experiences a marked reduction to below $2\%$, while the ACC declines to approximately $1\%$ before and after implementing mitigation. 

In the case of patch-based SSL-Backdoor, our mitigation approach significantly reduces the ASR to below $1.2\%$, demonstrating its successful eradication of backdoor effects.
Additionally, the ACC experiences an average decrease of $1.15\%$ in backdoored models after applying our mitigator. When utilizing the BYOL training method on the ImageNet-100 dataset, ACC remains relatively stable. 

Regarding global frequency-based SSL attack CTRL, our mitigation strategy substantially decreases the ASR to less than $2\%$, underscoring its effective elimination of backdoor attacks. Furthermore, the ACC has an average decline of $1.06\%$ in backdoored models after mitigation. Importantly, when employing the BYOL training technique on both CIFAR-10 and ImageNet-100 datasets, ACC remains fairly consistent. 

\noindent\textbf{Efficiency on Downstream Tasks.} SSL-Cleanse demonstrates effective generalization across various downstream tasks.  As indicated in Table~\ref{t:different-downstreams}, SSL-Cleanse reliably brings the ASR below $2\%$ for different downstream tasks such as CIFAR-10, STL-10, and GTSRB datasets, when the SSL encoders were backdoored on CIFAR-10 dataset. In addition, SSL-Cleanse works well across all targeted task labels. For CIFAR-10, it is possible to introduce a backdoor into any label from 0 to 9. SSL-Cleanse has been tested and proven to identify attacks on each label with a detection accuracy (DACC) exceeding $75\%$, and the average DACC across 10 labels is  $\sim 80\%$.

\begin{table}[h!]
\vspace{-0.15in}
\centering
\scriptsize
\setlength{\tabcolsep}{4pt}
\caption{SSL-Cleanse performance on different downstream tasks.}
\begin{tabular}{ccccc}\toprule
\multirow{2}{*}{Downstream}  & \multicolumn{2}{c}{Before Mitigation} & \multicolumn{2}{c}{After Mitigation} \\ \cmidrule(lr){2-3} \cmidrule(lr){4-5}
Task &  ACC(\%) & ASR(\%) & ACC(\%) & ASR(\%)\\\midrule
CIFAR-10   & $83.42$ & $48.32$ & $82.14$ & $1.14$ \\
STL-10  & $77.53$ & $36.55$ & $75.04$ & $1.84$ \\
GTSRB  & $72.01$ & $38.28$ & $70.98$ & $1.03$\\\bottomrule
\end{tabular}
\label{t:different-downstreams}
\vspace{-0.15in}
\end{table}

\begin{wrapfigure}{r}{0.5\textwidth}
\vspace{-0.3in}
  \centering
  \includegraphics[width=0.5\textwidth]{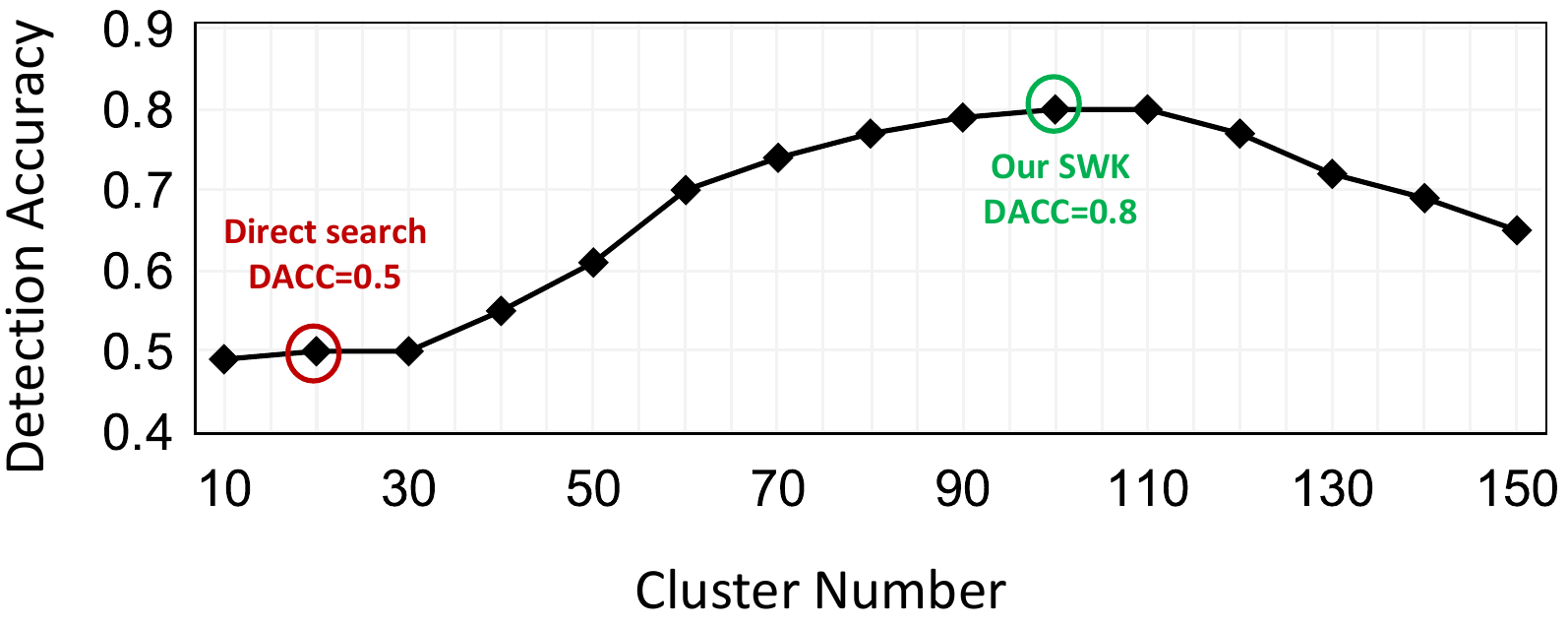} 
  \caption{A comparison of detection accuracy between SSL-Cleanse using the SWK method and the direct search (Kneedle)on ImageNet-100.}
 \label{fig:DetectionACC}
 \vspace{-0.25in}
\end{wrapfigure}

\noindent\textbf{The SWK Effects.} Figure~\ref{fig:DetectionACC} presents an ablation study on our SWK approach. Employing SWK on ImageNet-100 yields a predicted number of \(100\), in contrast to \(20\) as obtained using the direct search (Kneedle) method, as depicted in Figure~\ref{fig:SWK}. Consequently, SWK delivers a detection accuracy of \(80\%\), marking a \(30\%\) DACC improvement.

\noindent\textbf{Data Ratio Effects.} We analyze the influence of different  dataset ratios on detection performance using CTRL attack and BYOL training methods. We  employ varied image proportions, namely, $5\%$, $8\%$, and $10\%$. As illustrated in Table \ref{tab:data_ratio}, the results showcase an improvement in detection performance as the proportion of the training dataset increases. The DACC for the CIFAR-10 dataset shows a $1\%$ increase between the $8\%$ and $10\%$ training datasets. Moreover, for the ImageNet-100 dataset, there is an enhancement from $83\%$ to $85\%$ in the DACC. Further details regarding the hyperparameters \(\lambda\), along with nuances about the influence of attacking trigger size and perturbation norm, can be found in our supplementary material.

\begin{table}[h!]
\vspace{-0.15in}
\caption{Influence of the ratio of the unlabeled dataset to the entire training dataset on the detection efficacy. A larger ratio usually introduces a higher DACC. }
\vspace{-0.1in}
\scriptsize
\setlength{\tabcolsep}{4pt}
\begin{center}
\begin{tabular}{ccccccc}\toprule
\multirow{2}{*}{\makecell[c]{Data ratio\\(\%)}} & \multicolumn{3}{c}{CIFAR-10} & \multicolumn{3}{c}{ImageNet-100} \\
\cmidrule(lr){2-4}\cmidrule(lr){5-7}
& TP & FP & DACC(\%) & TP & FP & DACC(\%) \\\midrule
$5$ & $28$ & $7$ & $71$ & $38$ & $6$ & $82$ \\
$8$ & $37$ & $8$ & $79$ & $40$ & $7$ & $83$ \\
$10$ & $38$ & $8$ & $80$ & $43$ & $8$ & $85$ \\
\bottomrule
\end{tabular}
\label{tab:data_ratio}
\vspace{-0.15in}
\end{center}
\end{table}

\begin{figure}[h!]
\vspace{-0.2in}
\centering
\includegraphics[width=0.7\linewidth]{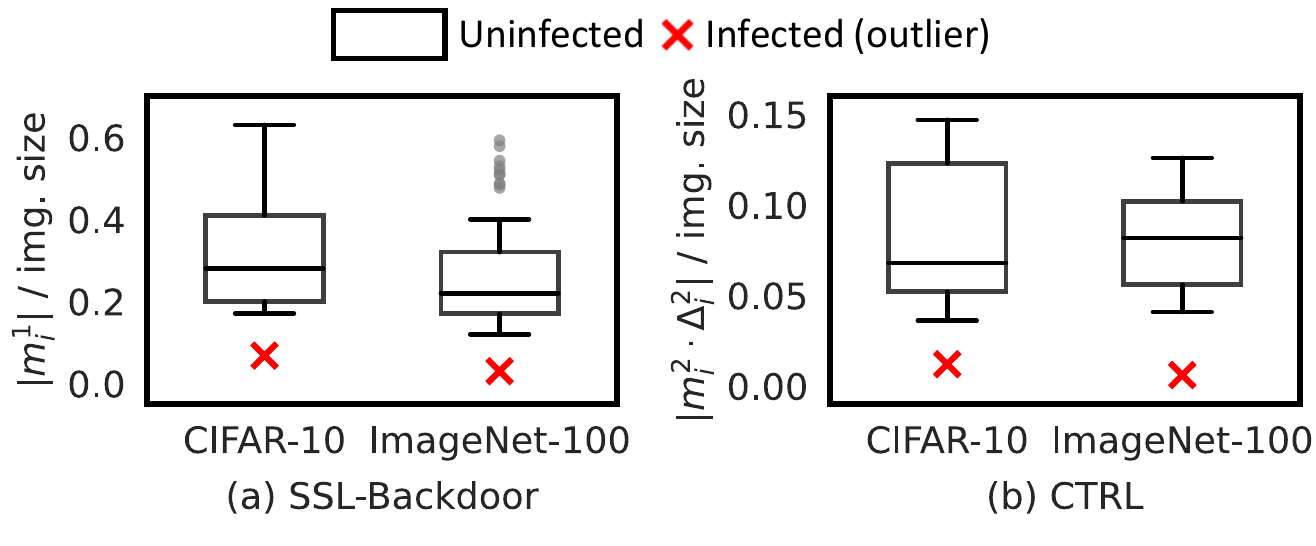}
\vspace{-0.1in}
\caption{The Size-Norm trigger outlier detection criteria is able to identify both patch-based SSL-Backdoor and frequency-domain global trigger in CTRL. The box plot shows min/max and quartiles.}
\label{fig:box2}
\vspace{-0.2in}
\end{figure}

\noindent\textbf{Outlier Detection Effects.} Figure~\ref{fig:box2} shows our Size-Norm Trigger Outlier Detector can successfully distinguish the outlier triggers from the $K$-cluster trigger list for the trojaned encoders based on SSL-Backdoor and CTRL. In particular, each box bar plots the trigger size $|m_i^1|$ and trigger norm $|m_i^2\cdot \Delta_i^2|$ for SSL-Backdoor and CTRL, respectively. Our Size-Norm trigger outlier detection can scan collectively, without the presumption that the defender has prior knowledge of the attack type.

\section{Conclusion} \label{conclusion}
This paper introduces \textit{SSL-Cleanse}, a novel work to detect and mitigate Trojan attacks in SSL encoders without accessing any downstream labels. We evaluated SSL-Cleanse on various datasets using 1200 models, achieving an average detection success rate of $82.2\%$ on ImageNet-100. After mitigating backdoors, backdoored encoders achieve $0.3\%$ average attack success rate without great accuracy loss.

\newpage

\bibliographystyle{splncs04}
\bibliography{main}
\end{document}